\documentclass[preprint,12pt]{elsarticle}
\usepackage{setspace}
\usepackage[]{natbib}
\usepackage{graphicx,amssymb,amsmath,times}
\usepackage{subfigure}
\usepackage{lscape}
\journal{New Astronomy}
\begin{document}
\begin{frontmatter}
\title{A time dependent approach to model X-ray and $\gamma$--ray light curves of Mrk 421 observed during the flare 
       in February 2010}
\author[label1]{K K Singh\corref{corr}}
\ead{kksastro@barc.gov.in}
\author[label1]{S Sahayanathan} \author[label2]{A Sinha} \author[label1]{N Bhatt} \author[label1]{A K Tickoo} 
\author[label1]{K K Yadav} \author[label1]{R C Rannot}  \author[label1]{P Chandra} \author[label1]{K Venugopal}
\author[label1]{P Marandi} \author[label1]{N Kumar} \author[label1]{H C Goyal} \author[label1]{A Goyal} 
\author[label1]{N K Agarwal} \author[label1]{M Kothari} \author[label1]{K Chanchalani} \author[label1]{V K Dhar} 
\author[label1]{N Chouhan} \author[label1]{C K Bhat} \author[label1]{M K Koul} \author[label1]{R Koul} 
\address[label1]{Astrophysical Sciences Division, Bhabha Atomic Research Center. \\
Mumbai- 400 085, India.}
\address[label2]{Department of High Energy Physics, Tata Institute of Fundamental Research, Homi Bhabha Road.\\
Mumbai- 400 005, India}
%---------------------------------------Abstract----------------------------------------------------------
\begin{abstract}
Strong X-ray and $\gamma$--ray flares have been detected in February 2010 from the high synchrotron peaked blazar 
Mrk 421 (z=0.031). With the motivation of understanding the physics involved in this flaring activity, we study  
the variability of the source in X-ray and $\gamma$--ray energy bands during the period February 10-23, 2010 
(MJD 55237-55250). We use near simultaneous X-ray data collected by \emph{MAXI}, \emph{Swift}-XRT and $\gamma$--ray 
data collected by \emph{Fermi}-LAT and \emph{TACTIC} along with the optical V-band observations by \emph{SPOL} 
at Steward Observatory. We observe that the variation in the one day averaged flux from the source during the flare 
is characterized by fast rise and slow decay. Besides, the TeV $\gamma$--ray flux shows a strong correlation with 
the X-ray flux, suggesting the former to be an outcome of synchrotron self Compton emission process. To model the observed 
X-ray and $\gamma$--ray light curves, we numerically solve the kinetic equation describing the evolution of particle 
distribution in the emission region. The injection of particle distribution into the emission region, from the putative 
acceleration region, is assumed to be a time dependent power law. The synchrotron and synchrotron self Compton emission 
from the evolving particle distribution in the emission region are used to reproduce the X-ray and $\gamma$--ray flares 
successfully. Our study suggests that the flaring activity of Mrk 421 can be an outcome of an efficient acceleration 
process associated with the increase in underlying non-thermal particle distribution. 
\end{abstract}
%---------------------------------------Keywords-------------------------------------------------------------

\end{frontmatter}

%---------------------------------------Section-1:Introduction------------------------------------------------
\section{Introduction}  
Blazars belong to the extreme class of radio-loud active galactic nuclei (AGN) hosting a relativistic jet which is pointed 
close to the line of sight from the Earth. The blazar emission originating from the jet is observed to be highly variable 
over all accessible energy bands. The radiation from blazars is dominated by non-thermal emission that spans over the entire 
electromagnetic spectrum from radio to very high energy (VHE, E$>$100 GeV) $\gamma$--rays. The multi-wavelength radiation 
observed from blazars is best understood in a paradigm where the emission results from the magnetized plasma ejected with 
relativistic speeds in a collimated outflow powered by a super massive black hole. The spectral energy distribution (SED) 
of blazars is characterized by two distinct and broad peaks located in the UV/optical/X-ray and in high energy 
(HE, E$>$100 MeV) $\gamma$--ray regimes respectively. The low energy component is believed to be originated due to 
synchrotron radiation from a population of relativistic electrons losing energy in the jet magnetic field. However, the origin 
of second component peaking  in HE $\gamma$--ray regime, is poorly understood and is still under debate. In leptonic models, 
this emission originates from the inverse Compton (IC) scattering of ambient photons by the same electrons that emit the 
synchrotron radiation \cite{Maraschi1992, Botcher2007}. The ambient photon field can be synchrotron radiation in the emission 
region or the photons external to the jet. In synchrotron self Compton (SSC) model, only local synchrotron photons serve as 
seed photons for IC scattering \cite{Maraschi1992, Tavecchio1998}; whereas, in external Compton (EC) models, the target photon 
fields can be direct accretion disk field \cite{Dermer1993}, the line emission from the broad line emitting region and IR 
radiation from dusty torus \cite{Sikora1994}. On the other hand, in hadronic models, hadronic processes such as proton 
synchrotron and radiation produced by secondary particles are suggested to explain the emission of HE and VHE $\gamma$--ray 
radiation from blazars \cite{Mannheim1993, Mucke2003}.
\par
Although, the broad band spectral properties are nearly similar in all blazars, they are further classified into two main 
subclasses namely BL Lacertae objects (BL Lacs) and flat spectrum radio quasars (FSRQs), on the basis of their optical 
line features \cite{Urry1995}. FSRQs display strong, broad optical absorption/emission lines while BL Lacs are characterized by 
featureless continuum showing weak/no emission/absorption lines. According to the \emph{blazar simplified view}, BL Lacs and 
FSRQs are explained as low excitation radio galaxies (Fanaroff-Riley I) and high excitation (Fanaroff-Riley II) radio galaxies 
with their jets forming a small angle with respect to the line of sight \cite{Urry1995, Giommi2012}. This scenario is consistent 
with the complex observational properties of blazars from the surveys carried out so far in the radio and X-ray energy bands.
Besides this, the characteristic peak frequencies of synchrotron and IC components of blazar SED are anti-correlated with 
respect to bolometric luminosity \cite{Fossati1998, Ghisellini2016}, leading to further classification of blazars. Based on the peak 
frequency ($\nu_p^{sync}$) of the synchrotron component, blazars are divided into three classes: low-synchrotron peaked 
(LSP: $\nu_p^{sync} < 10^{14}$ Hz), intermediate-synchrotron peaked (ISP: $10^{14} < \nu_p^{sync} < 10^{15}$ Hz) and high-synchrotron 
peaked (HSP: $\nu_p^{sync} > 10^{15}$ Hz) blazars \cite{Abdo2010}. Almost all FSRQs detected so far belong to the LSP 
class of blazars whereas BL Lacs include LSP, ISP and HSP blazars. The ratio of peak frequencies of X-ray and $\gamma$--ray 
components gives an idea about the characteristic cooling energy of relativistic particles in the jet plasma under the assumption 
that the emission region is homogeneous in these energy regimes. Recent multi-wavelength observations of 42 radio-loud narrow-line 
Seyfert 1 (RLNLS1) galaxies have revealed similarities between blazars and these peculiar systems \cite{Foschini2015}. Detection of 
$\gamma$--ray emission from RLNLS1 galaxies have strong implications for the presence of blazar-like jets in these objects 
\cite{Abdo2009}. The main differences between blazars and RLNLS1s are in the masses of central engines and accretion rates. 
The blazars generally have larger masses and lower accretion rates whereas $\gamma$--ray emitting RLNLS1s have small masses and high 
accretion rates \cite{Yuan2008}. Although RLNLS1s display some peculiar differences with respect to blazars, the physical properties 
inferred from observations indicate that these sources are low mass tail of blazars. Therefore, blazars and RLNLS1s can be 
considered as similar objects and the differences in their observed SEDs are attributed to thier masses and accretion rates 
under the assumption that relativistic jets are formed independently of the host galaxies \cite{Abdo2009}. 
\par
The multi-wavelength emission from blazars is also characterized by rapid variability on various time scales ranging from few 
minutes to years. Based on these time scales, the variability of blazars is classified in three groups namely intra day, short 
time and long time. The intra day variability (IDV) implies time scales from few tens of minutes to less than a day 
\cite{Wagner1995} and it is also known as micro or intra night variability. The short time variability (STV) is characterized 
by time scales from several days to few months whereas long time variability (LTV) covers changes from several months to many 
years \cite{Gupta2004}. The shortest variability time scales from blazar light curves can be used to constrain the size of the
emission region through light travel time effects, which inturn probes the location of the emission zone in the self-similar jets. 
\par
Mrk 421 (z=0.031, 134 Mpc), an HSP type BL Lac object, was first detected at energies above 100 MeV by EGRET telescope on the Compton 
Observatory \cite{Lin1992}. The source is also the first extragalactic object observed by ground based imaging atmospheric Cherenkov 
telescope (IACT) at energies in the VHE regime \cite{Punch1992}. Mrk 421 has been observed to be a very active blazar with major outbursts 
about once every year in both X-ray and $\gamma$--ray energy bands with variability time scales from minutes to several days 
\cite{Cui2004, Tluczykont2010}. In February 2010, this source was detected in a state of high activity reaching its maximum 
around February 16, 2010. During this outburst, the source was the target of a multi-wavelength campaign involving X-ray satellites, 
$\gamma$--ray satellites, ground based VHE instruments, optical and radio telescopes. Near simultaneous  multi-wavelength data from 
radio to TeV $\gamma$--rays collected during the campaign have been used to perform an extensive study of the source during this 
flaring activity \cite{Shukla2012, Singh2015}. X-ray and HE $\gamma$--ray observations of the same flare with \emph{Swift} and 
\emph{Fermi} satellites have also been studied separately \cite{Singh2012}. Two X-ray flares from Mrk 421 detected around the same 
period with \emph{MAXI} satellite during January and February 2010 are also reported \cite{Isobe2010, Gaur2012}. The VHE observations 
of the source during this flare have also been reported by various IACT groups like VERITAS \cite{Fortson2012}, HESS \cite{Tluczykont2011} 
and TACTIC \cite{Singh2015}. Long term VHE observations of the source during 2009--10 with TACTIC have also been reported \cite{Chandra2012}.
\par
Recently, Zheng et al.(2014) have studied the February 2010 outburst of Mrk 421 considering stochastic acceleration of electrons 
under strong magnetic turbulences \cite{Zheng2014}. They studied the flare assuming different indices of the turbulent spectrum and 
conclude that the observed properties of the flare favour the turbulent index to be $\sim$ 2 (hard sphere approximation). 
Yan et al.(2013) have investigated the electron energy distribution and acceleration processes in the jet of Mrk 421 through 
SED fitting in different active states under the framework of one zone SSC model \cite{Yan2013}. They assume two electron 
energy distributions formed via different acceleration processes: the shock accelerated power law with exponential cut-off and 
the stochastic turbulence accelerated log-parabolic distributions to model the SED of the source. Their results indicate that 
both the electron energy distributions give good fits for the SED in low state, but log-parabolic distribution gives better 
fits for the SED in high or flaring state and the VHE variability can be accomodated only in log-parabolic model. 
The X-ray and TeV variability of Mrk 421 flare observed during March 2001 have been investigated by \cite{Mastichiadis2013} 
in leptohadronic single zone model. They find that $\gamma$--ray emission can be attributed to the synchrotron radiation either 
from protons or from secondary leptons produced via photohadronic processes. These possibilities also imply differences in 
the variability signatures in X-ray and $\gamma$--ray regime. Chen et al. (2011) have reported a time dependent simulation 
of multi-wavelength variability of Mrk 421 with a Monte Carlo multizone radiative transfer code \cite{Chen2011}. The variability 
in this code is introduced by the injection of relativistic electrons as a shock front crosses the emission region. The emission 
is considered to originate from two components: a flaring or active and a quasi-steady state. The later one is completely independent 
of the former component. 
\par
In the present work, we use well sampled X-ray and $\gamma$--ray daily light curves to investigate the temporal properties of Mrk 421 
during the flare in February 2010. We show that a time dependent power law injection of electrons in the emission region can be used to 
understand the outburst of the source in different energy bands within the framework of SSC model. The paper is organized as follows: 
Section 2 presents the details of Optical, X-ray and $\gamma$--ray data used in this study. In Section 3, we perform the temporal and 
correlation study of the daily light curves. Modelling of X-ray and $\gamma$--ray light curves using time dependent injection of particles 
is described in Section 4. In Section 5, we discuss and conclude our results. $\Lambda$CDM cosmology with $\Omega_m = 0.3$, 
$\Omega_\Lambda = 0.7$ and $H_0 = 70\;km\;s^{-1}\;Mpc^{-1}$ is adapted in the present work.

%------------------------------Section- 2:MWL Data Set------------------------------------------------------------------------------------
\section{Multi-wavelength Data Set}
In February 2010, Mrk 421 went into a state of high activity with maximum flux detected around February 16, 2010 (MJD 55243) in X-ray 
and $\gamma$--ray energy bands. Many instruments world wide have detected this outburst. The near simultaneous multi-wavelength observations 
during the period February 10-23, 2010 (MJD 55237-55250) have been used in the present study. We have obtained the V-band optical data from the 
\emph{SPOL} observations at Steward Observatory, University of Arizona\footnote{{http://james.as.arizona.edu/~psmith/Fermi/}} under 
\emph{Fermi} multi-wavelength blazar observing support program. 
The soft X-ray data in the energy range 0.3-10 keV from X-ray Telescope (XRT) onboard the \emph{Swift} satellite \cite{Burrows2005}
have been processed with the \emph{XRTDAS} software package (v.3.0.0) available within \emph{HEASOFT} package (6.16). We have used standard 
procedures (\emph{xrtpipeline v.0.13.0}) for cleaning and calibration of event files with data taken in Window Timing (WT) mode. 
The spectra are grouped using \emph{grppha v.3.0.1} to ensure a minimum of 30 counts in each bin. The daily light curves in the 
energy range 0.3-10keV are obtained using \emph{xrtproducts v.0.4.2}. 
The archival X-ray data in the energy range 10-20 keV during the same period from \emph{Monitor of All sky X-ray Image 
(MAXI)} satellite \cite{Matsuoka2009} have also been used in this study. We have obtained the daily light curve from \emph{MAXI} available 
online from its website\footnote{http://maxi.riken.jp/top/index.php}.  
\par 
The $\gamma$--ray observations in HE and VHE bands have been used from \emph{Fermi}-Large Area Telescope (LAT) \cite{Atwood2009} and 
TeV Atmospheric Cherenkov Telescope with Imaging Camera \emph{(TACTIC)} \cite{Koul2007} observations of Mrk 421 during February 10-23, 2010 
respectively. The HE measurements from \emph{Fermi}-LAT over the energy range 0.1-100 GeV are obtained by analysing the data using standard 
\emph{Fermi}-ScienceTools. The details of data analysis procedure for generating the daily light curve in HE band with \emph{Fermi}-LAT are 
discussed in our earlier work \cite{Singh2015}. The one day averaged VHE light curve in the energy range 1.1-12 TeV is obtained from 
\emph{TACTIC} observations of the source during the above mentioned period. The procedure for analysis of \emph{TACTIC} data used in this 
work has been described in detail in \cite{Singh2015, Chandra2012}. The VHE flux points have been corrected for extragalactic background 
light (EBL) absoprtion of TeV photons using the model proposed by \cite{Franceschini2008}. The intrinsic VHE light curve obtained with 
\emph{TACTIC} indicates a strong flaring activity on February 16, 2010 (MJD 55243) which is also observed in X-ray and HE $\gamma$--ray bands.
%-----------------------------Section-3: Temporal and Correlation Analysis-----------------------------------------------
\section{Temporal Study}
\begin{figure}
\begin{center}
\includegraphics[width=1.0\textwidth]{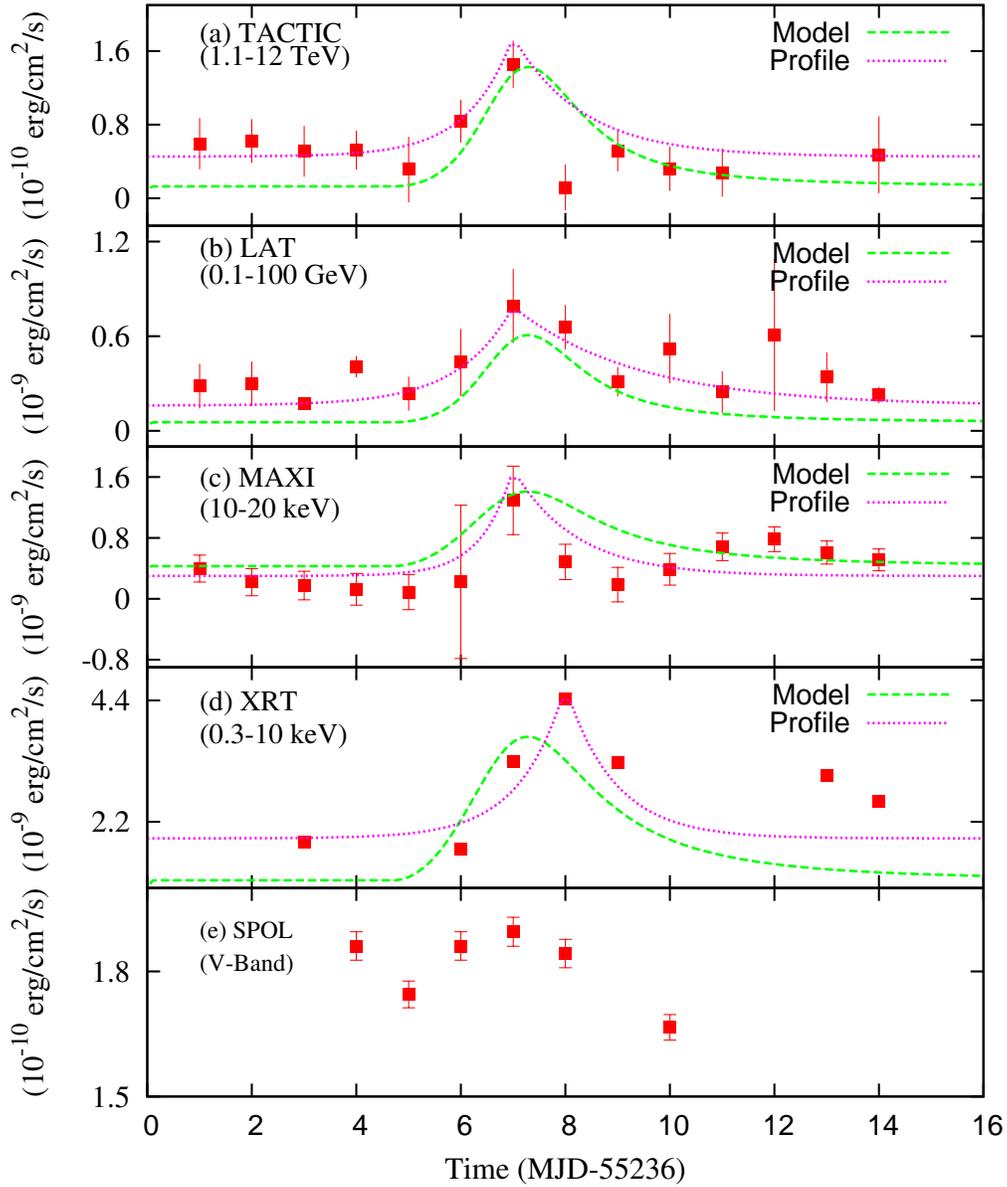}
\caption{ $\gamma$--ray, X-ray and Optical light curves of Mrk 421  during February 10-23, 2010 covering VHE $\gamma$--rays by 
\emph{TACTIC}, HE $\gamma$--rays by \emph{Fermi}-LAT, X--rays by \emph{MAXI} and \emph{Swift}-XRT and V-band optical by \emph{SPOL}. 
The dotted (magenta) line represents the best-fit of flare profile (equation 1) to the light curve in each panel, while the 
dashed (green) line shows the flux evolution in one zone SSC model with time dependent injection.}
\label{fig:Fig1}
\end{center}
\end{figure}

The resulting $\gamma$--ray, X-ray and Optical daily averaged light curves observed by different instruments during February 10-23, 2010 
(MJD 55237-55250) are shown in Figure \ref{fig:Fig1}(a-e). A significant enhancement of the flux in all energy bands is observed on  
February 16, 2010 (MJD 55243). To study the nature of this flare in detail, we fit the light curves with an exponential profile defined by 
\begin{equation} 
	F(t)=A_0 + A\left[e^{(t-t_p)/\tau_r} H(t_p-t) + e^{-(t-t_p)/\tau_d } H(t-t_p) \right]
\end{equation}
where $\tau_r$ and $\tau_d$ are the rise and decay timescales respectively, $t_p$ is the time of maximum flux, $A_0$ and $A$ are the constants 
for determining the quiescent and peak fluxes respectively, and $H$ is the \emph{Heaviside function}. The fitting is performed in 
X-ray and  $\gamma$--ray energy bands over the parameters $\tau_r$, $\tau_d$, $A_0$ and $A$, whereas $t_p$ is fixed at 55243 MJD. 
The best fit values of these parameters for different energy bands are presented in Table \ref{tab:Table1}. 
\par
The peak flux point observed with \emph{Swift}-XRT appears delayed by about one day with respect to \emph{MAXI}, \emph{LAT} and \emph{TACTIC} 
bservations. It is important to note here that the XRT data is of very short duration (less than an hour) while the data from other instruments 
are averaged over a day. Also \emph{TACTIC} observations show a sharp drop just after the flare, which is not observed in other energy bands. 
Almost similar rise times in all the light curves suggest that the physical process involved in the flare is energy independent. Likewise, 
almost similar decay times in all energy bands do not favour the flare decay due to radiative loss mechanism \cite{Bottcher2009}. 
The fitted light curves for four energy bands are shown in Figure \ref{fig:Fig1}(a-d) as dotted (magenta) lines. It is quite evident from the 
Figure \ref{fig:Fig1}(a-d) that the flare is asymmetric with fast rise and slow decay. We observe that the optical light curve with a high 
flux point on February 13, 2010 (MJD 55240) is different from the light curves in other energy bands. In addition, the features in optical 
light curve are not coincident with those in X-ray and $\gamma$--ray light curves because the optical emission appears to be in high state 
even before the beginning of the flare. These differences suggest that the optical emission may not be associated with the region responsible 
for emission in higher energy bands \cite{Aleksic2015, Atreyee2013}. This claim is further strengthened by the spectral behaviour of the 
broadband SED during flare maximum (Figure \ref{fig:Fig2}) where, the synchrotron spectral component significantly underpredicts the observed optical 
flux. Due to these reasons, we have not included optical emission in our present study.
%------------------------------------------Figure 1----------------------------------------------------------------------------
%--------------------------------Table-1 Fitted Parameters of Temporal Analysis------------------------------------------------
\begin{table}
\caption{Best-fit parameters of the temporal profile (equation 1) fitted to the light curves in X-ray, GeV and TeV energy ranges. 
The corresponding fitted profiles alongwith the data are shown in Figure \ref{fig:Fig1}(a-d)}.
\begin{center}
\begin{tabular}{cccccc}
\hline
Energy  	&A$_0$	&A	&$\tau_r$	&$\tau_d$	&$\chi^2_r$(dof)\\
		&(erg cm$^{-2}$ s$^{-1}$)&(erg cm$^{-2}$ s$^{-1}$) &(days) &(days)	&\\
\hline
1.1-12 TeV	&(4.53$\pm$1.52)$\times10^{-11}$ &(1.30$\pm$0.32)$\times10^{-10}$&(0.86$\pm$0.38)&(1.32$\pm$0.48)&1.46(8)\\
0.1-100 GeV     &(1.61$\pm$0.39)$\times10^{-10}$ &(6.28$\pm$2.21)$\times10^{-10}$&(1.04$\pm$0.80)&(2.37$\pm$1.32)&1.58(10)\\
10-20 keV	&(3.12$\pm$0.94)$\times10^{-10}$ &(1.40$\pm$0.57)$\times10^{-09}$&(0.60$\pm$0.31)&(1.22$\pm$0.42)&1.82(10)\\
0.3-10 keV	&(1.90$\pm$0.25)$\times10^{-09}$ &(2.74$\pm$1.34)$\times10^{-09}$&(0.87$\pm$0.72)&(0.94$\pm$0.85)&5.64(3)\\
\hline
\end{tabular}
\end{center}
\label{tab:Table1}
\end{table}

%-----------------------------------------------Section-4 Physical modeling of light curves -----------------------------   
\section{Physical Modeling of Light curves}
To obtain the physical parameters governing the broadband emission from the jet of Mrk 421, we assume the emission to arise 
from a spherical region of radius $R$ moving down the jet with Lorentz factor $\Gamma$ at viewing angle $\theta$ with respect 
to the line of sight of the observer. The low energy emission from radio to X--rays is interpreted as synchrotron radiation from 
a power law distribution of non-thermal particles, $N(\gamma)d\gamma=K\gamma^{-p}d\gamma$, losing their energy ($\gamma m c^2$) in 
a tangled magnetic field $B$ and the high energy emission in $\gamma$--rays is attributed to the SSC scattering process 
\cite{Sahayanathan2012}. The observed light curves of Mrk 421 during flaring activity are modelled using the evolution of  
spectrum from the emission region. The evolution of particle distribution $n(\gamma, t)$ in the emission region can be described 
by the kinetic equation     
\begin{equation}\label{eqn:kinetic}
	\frac{\partial}{\partial t} n(\gamma, t) = \frac{\partial}{\partial \gamma}[P(\gamma, t)  n(\gamma, t)] - \frac{n(\gamma, t)}
                                                  {t_{esc}} + Q(\gamma, t)
\end{equation}
where $P(\gamma, t)$ is the energy loss rate and $Q(\gamma, t)$ is the particle injection rate from an associated acceleration region. 
We consider the energy loss rate to be dominated by the synchrotron and inverse Compton emission processes and for the latter we use 
the exact Klein Nishina cross section \cite{Rybicki1986, Blumenthal1970}. The injection spectrum from the acceleration region is 
assumed to be a time dependent power law of the form  
\begin{equation}
	Q(\gamma, t)= K(t)\gamma^{-p}	~~~~~~~;~~~\gamma_{min} < \gamma < \gamma_{max}
\end{equation}
with a time dependent normalization $K(t)$ given by 
\begin{equation}
	K(t)=K_0 + 2(K_p - K_0)\left[\left(\frac{t-t_0}{t_p-t_0}\right)^{-\alpha_1} + 
                               \left(\frac{t-t_0}{t_p-t_0}\right)^{\alpha_2}\right]^{-1}
\end{equation}
where $K_0=K(t_0)$ is the normalization in quiescent state and $t_p$ is the time when flare attains the peak with $K_p=K(t_p)$ as the 
corresponding normalization, and $\alpha_1$ and $\alpha_2$ decide the rise and decay time scales of the particle injection. The time 
dependence in the particle injection will result in the inverse Compton target photon density $U_{syn}(\gamma, t)$ to vary with time and 
this will lead to time dependent energy loss rate, preventing equation \ref{eqn:kinetic} to be solved analytically. Hence, we solve 
equation \ref{eqn:kinetic} numerically using finite difference method \cite{Chiaberge1999, Chang1970, Press2002}. The emission due to 
synchrotron and SSC processes are obtained by convoluting the single particle emissivity with the evolving particle distribution from 
the emission region. Finally, the flux received by the observer at frequency $\nu_{obs}$ at time $t_{obs}$ is transformed considering 
the relativistic Doppler boosting and cosmological effects \citep{Begelman1984}
\begin{equation}
	F_{obs}(\nu_{obs}, t_{obs})=\frac{\delta^3(1+z)}{d_L^2}V j(\nu, t) 
\end{equation}
Here $d_L$ is the luminosity distance, $V$ is the comoving volume of emission region and $j(\nu, t)$ is the emissivity of source at 
frequency $\nu=(\frac{1+z}{\delta})\nu_{obs}$ at time $t=(\frac{\delta}{1+z})t_{obs}$ due to different radiative processes. 
\par
\begin{figure}
\begin{center}
\includegraphics[width=0.70\textwidth,angle=-90]{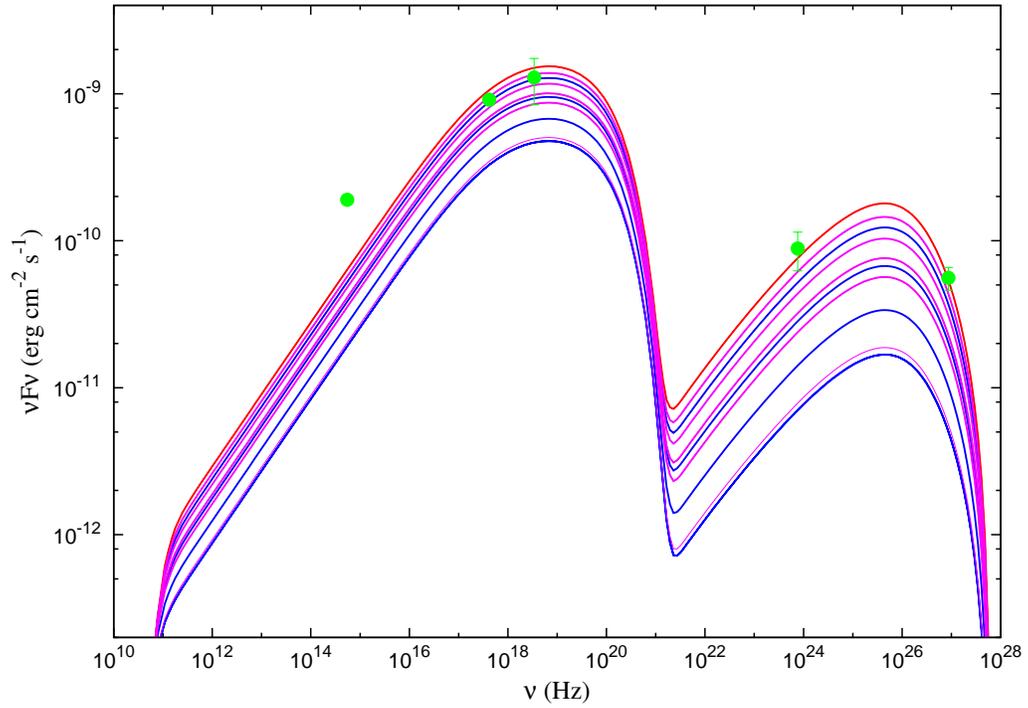}
\caption{Evolution of SED of Mrk 421 during the flare under the framework of time dependent SSC model. Red line shows the model SED 
at the peak of the flare. Model SEDs shown by blue lines represent the time evolution during the rise of the flare while SEDs in 
magenta show time evolution during the fall of the flare. The measured flux points in different energy bands correspond to the 
values observed during the flare on February 16, 2010. The VHE flux point measured with \emph{TACTIC} in the energy range 1.1-12 TeV 
is corrected for EBL absorption using the model proposed in \cite{Franceschini2008}.}
\label{fig:Fig2}
\end{center}
\end{figure}

We apply this time dependent model on the X-ray and $\gamma$--ray light curves of Mrk 421 observed during its flare on February 
16, 2010 (MJD 55243). The energy integrated photon flux obtained from the time dependent particle spectrum is used to reproduce 
the light curves in four energy bands : 0.3-10 keV, 10-20 keV, 0.1-100 GeV and 1.1-12 TeV. The parameters of the model are further 
optimized through the multi-wavelength spectral fitting of time averaged SED observed during the flare of the source. The evolution 
of SED of the source during the flare is shown in Figure \ref{fig:Fig2}. The escape time scale is considered to be $t_{esc}=0.1R/c$. 
The particle injection into the emission region is initiated on MJD 55240 and the time of maximum particle flux injection, $t_p$ is 
fixed at MJD 55243. We find that increasing the electron injection three times compared to the quiescent level can reproduce the X-ray 
and $\gamma$--ray light curves satisfactorily. The model light curves in four energy bands (lines with larger dots) along with the 
observed daily flux points are shown in Figure \ref{fig:Fig1}. The source parameters describing the model light curves are summarized in 
Table \ref{tab:Table2} and are consistent with the values obtained from the SED modeling \cite{Singh2015, Acciari2011}.
%----------------------------------------Figure-2 SED----------------------------------------------------------
%--------------------------------Table-3 SED Fitted Parameters---------------------------------
\begin{table}
\caption{Parameters of time dependent SSC model from X-ray and $\gamma$--ray light curves fitting.}
\begin{center}
\begin{tabular}{lclc}
\hline
Parameters  		&Value \\
\hline
R (cm)			&8.0$\times 10^{15}$ \\
$\delta$                &18\\
p			&2.02\\
B (G)			&0.60\\
$\gamma_{min}$		&12\\
$\gamma_{max}$	        &3$\times10^6$\\
$K_0$$(cm^{-3} s^{-1})$	&3.0$\times 10^{-2}$ \\
$K_p$$(cm^{-3} s^{-1})$	&1.0$\times 10^{-1}$ \\
$\alpha_1$		&2.1\\
$\alpha_2$ 		&3.2\\
\hline
\end{tabular}
\end{center}
\label{tab:Table2}
\end{table}
%---------------------------------------------------Correlation-----------------------------------------------------------
\par
In addition to the modeling of light curves, we also perform a correlation study between $\gamma$--ray and X-ray energy bands for the emission 
of the source, since this would provide useful information regarding the processes responsible for the flaring activity. 
To investigate the possible correlation, we follow the similar procedure as proposed by  Katarzy{\'n}ski et al. 2010 \cite{Katarzynski2010}. 
We find that the correlation between the evolution of  emissions in the two different energy regimes during the flare of Mrk 421 can be described by 
a simple power law of the form
\begin{equation} 
	F_1 (t) \propto F^{\rho}_2 (t)
\end{equation} 
where $\rho$ is the index of correlation for the near simultaneous flux measurements $F_1$ and $F_2$ in two independent energy bands. 
The values of index of correlation for comparision of observations with different instruments are given in Table \ref{tab:Table3}.
This suggests strong correlation of the TeV $\gamma$--ray flux with respect to X--ray flux observed with \emph{MAXI}, which can be 
understood within the frame work of SSC model. If the flaring activity is predominantly attributed to the variation in non-thermal 
particle density, the SSC flux ($\gamma$--ray) will vary quadratically with respect to the synchrotron flux (X-ray) due to the second 
order dependence of the former with the particle density \cite{Katarzynski2010}. A linear correlation may be an artifact of the decay of 
the flare when the emitting region expands and it leads to a decrease in electron density and magnetic field strength. 
%----------------------------------Table-2: Index of Correlation-------------------------------------------------------------------------------
\begin{table}
\caption{Index of cross correlation ($\rho$) for near simultaneous observations with different $\gamma$--ray and X-ray instruments.}
\begin{center}
\begin{tabular}{lclc}
\hline
Instruments  	&$\rho$\\		
\hline
TACTIC-MAXI     &1.23$\pm$0.72\\
LAT-MAXI	&0.65$\pm$0.25\\
\hline
\end{tabular}
\end{center}
\label{tab:Table3}
\end{table}

%-------------------------------------Section-5- Discussion and Conclusions------------------------------------------------
\section{Discussion and Conclusions}
In this work, we have performed a detailed temporal study of X-ray and $\gamma$--ray observations on Mrk 421 during the flare in 
February 2010 to understand the flaring behaviour of the source. We determine the rise and decay characteristic timescales in 
the daily light curves of the source in four energy bands: 0.3-10 keV, 10-20 keV, 0.1-100 GeV and 1.1-12 TeV. 
The flux points in light curves before and after the flare are fitted with exponential rise and decay profiles respectively. 
We obtain almost similar rise time in all the energy bands and this indicates an energy independent physical process operating in 
the source during the flaring activity. Also, similar characteristic decay times in different energy bands do not support the 
radiative loss processes involved in the decay of the flare. Based on these studies, we interpret the flare to be a result of 
enhanced relativistic particle distribution by a putative acceleration process and the emission in the $\gamma$--ray energy regime 
to be dominated by SSC process. In addition, a strong correlation observed between X-ray and TeV fluxes during the flare supports 
SSC interpretation of VHE emission.
\par
To further understand the behaviour of the source during the flare, the X-ray and $\gamma$--ray light curves are reproduced using 
a time dependent model. The evolution of particle distribution in the cooling region is obtained by solving the relevant kinetic 
equation considering synchrotron and SSC emission processes. The evolution of radiation flux can reproduce the observed daily 
light curves in X--rays and $\gamma$--rays satisfactorily. The reproduction is better for the $\gamma$--ray light curves than the 
X-ray ones though the uncertainties in the former are relatively large. The variability in optical V-band is not significant 
as compared to X-ray and $\gamma$--ray bands and small variations in the flux level may be due to contamination from the host galaxy. 
We correct the optical data for starlight contribution from the host galaxy using the methodology described in \cite{Nilsson2007}. 
The effect of contamination from host galaxy of Mrk 421 on V-magnitude in our present study is obtained to be negligible. The optical 
emission during the flare is probably not related to that in X-ray and $\gamma$--ray energy bands and it is consistent with the results 
from previous studies of Mrk 421 \cite{Aleksic2015, Atreyee2013}.  In our simplistic model, we attribute the entire flare to the varying 
particle injection into the emission region. However, this may not be strictly true in a realistic scenario where the flare can be associated 
with variation in other parameters as well, for example magnetic field, bulk Lorentz factor etc. Reproduction of the light curves can be 
improved by evolving other parameters along with the particle density. Nonetheless, this will lead to the increased number of parameters 
and a reasonable conclusion on the physical condition responsible for the flare cannot be derived. Again, we have not considered the light 
travel time effects \cite{Eichmann2010, Eichmann2012, Zacharias2013, Zacharias2014} since the flare time scale at the source 
frame ($t_{fl}^\prime$) is larger than the light travel time. For instance considering a flare time scale ($t_{fl}$) of $\sim$ 1 day in 
observer frame, the corresponding time in source frame will be $t_{fl}^\prime=\frac{\delta}{1+z}t_{fl}\approx$18 days; whereas the light 
travel time will be $R/c\approx$3 days. Hence, this effect will not be significant in our simple phenomenological model.
%---------------------------------------Acknowledgement------------------------------------------------------
\section*{Acknowledgment}
We are grateful to the anonymous referee for his/her valuable suggestions and comments. We would like to acknowledge the use of 
\emph{TACTIC} data and the excellent team work of colleagues in Astrophysical Sciences Division for their contribution to the 
simulation, observation, data analysis and instrumentation aspects of TACTIC gamma-ray telescope. We acknowledge the use of 
public data obtained through \emph{Fermi} Science Support Center (FSSC) provided by NASA. This work made use of data supplied 
by the UK Swift Science Data Centre at the University of Leicester. This research has made use of the MAXI data, 
provided by RIKEN, JAXA and the MAXI team. Data from the Steward Observatory spectropolarimetric monitoring project were used. This program is supported by Fermi Guest Investigator grants NNX08AW56G, NNX09AU10G, NNX12AO93G, and NNX15AU81G.
%-----------------------------------------References--------------------------------------------------------

\end{document}